# J-matrix method of scattering in one dimension:
## The nonrelativistic theory

A. D. Alhaidari[†], H. Bahlouli, and M. S. Abdelmonem

*Saudi Center for Theoretical Physics, Dhahran, Saudi Arabia*
AND
*Physics Department, King Fahd University of Petroleum & Minerals, Dhahran 31261, Saudi Arabia*

We formulate a theory of nonrelativistic scattering in one dimension based on the J-matrix method. The scattering potential is assumed to have a finite range such that it is well represented by its matrix elements in a finite subset of a basis that supports a tridiagonal matrix representation for the reference wave operator. Contrary to our expectation, the 1D formulation reveals a rich and highly non-trivial structure compared to the 3D formulation. Examples are given to demonstrate the utility and accuracy of the method. It is hoped that this formulation constitutes a viable alternative to the classical treatment of 1D scattering problem and that it will help unveil new and interesting applications.

## I. Introduction

One of the fundamental tools in physics that is used in probing atomic and subatomic phenomena involves the scattering of particles through a sample of the desired element. A typical setting of the scattering experiment in one dimension consists of a uniform beam of particles of known energy and current density incident on a target containing scattering centers. The incident beam of free particles is described by a plane wave of momentum *k* and unit amplitude. When this wave interacts with the scattering centers it results in a reflected and transmitted waves with reflection and transmission amplitudes that depend on the incident beam energy and properties of the target (parameters of the potential that models the structure of the target and its interaction with the incident beam).

We will be mainly concerned with the quantum scattering of particles constrained to move along a straight line (the *x*-axis) under the influence of a time independent real potential *V*(*x*). Such a one dimensional scattering is usually treated using the concepts of transmission and reflection coefficients. It has been shown in the past [1] that scattering in one dimension can be formulated in terms of phase shifts in close analogy to the three dimensional problem. As in three dimensions, it is reasonable to assume that at large distances from the scattering region the wavefunction can be written as a linear combination of plane waves [2]

$$\psi(x)=\begin{cases} e^{ikx}+Re^{-ikx} & ,x\to-\infty \\ Te^{ikx} & ,x\to+\infty \end{cases} \tag{1.1}$$

[†] Present temporary address: 1300 Midvale Ave. #307, Los Angeles, CA 90024

where $R$ and $T$ are the reflection and transmission amplitudes, respectively. The incident normalized wave is assumed to be injected from left and transmitted to the right. Probability conservation for hermitian Hamiltonians requires that $|R|^2+|T|^2=1$.

In this article, we introduce a formulation of the nonrelativistic scattering problem in one dimension based on the J-matrix method [3]. This is the first attempt at presenting a one-dimensional J-matrix formulation of scattering. As a result of this formulation, the usual thought-of simplicity of the 1D scattering problem when compared to 3D could be debated as evident in the non-trivial and rich structure emerging below. In this approach, we first deploy the analytic power inherent in the J-matrix method for the asymptotic solution of the reference Hamiltonian in which the theory of orthogonal polynomials plays an important role [4]. Then we exploit the efficiency of the numerical Gauss quadrature and continued fraction techniques [5] in the calculation of the reflection and transmission amplitudes. An interesting findings in this work is the proof that for any hermitian system with short-range symmetric potential, the transmission and reflection amplitudes can always be written in terms of phase factors, $T=\frac{1}{2}\left(e^{2i\theta_+}-e^{2i\theta_-}\right)$ and $R=\frac{1}{2}\left(e^{2i\theta_+}+e^{2i\theta_-}\right)$.

The J-matrix (Jacobi-matrix) theory of scattering is an algebraic method [3], which exploits the fact that the unperturbed (reference) Hamiltonian $H_0$ can be diagonalized in a certain complete set of square integrable ($L^2$) basis functions. The $L^2$ basis set is chosen such that the matrix representation of the reference Hamiltonian is tridiagonal. This requirement results in a three-term recurrence relation for the expansion coefficient of the wavefunction associated with $H_0$ and hence establishes a strong relationship between the solution of the reference problem and the corresponding orthogonal polynomials. This also allows for an exact analytic diagonalization of the reference Hamiltonian and expresses the wavefunction of the reference problem as an infinite convergent series with expansion coefficients that are written in terms of orthogonal polynomials. The scattering potential is assumed to be short-range and hence its representation can be truncated in a finite subset of the basis set. The accuracy of the computations can be improved and the errors can be made small by enlarging the basis set used in the truncated space where the potential is evaluated. Hence, the J-matrix differs from the algebraic variational theories in that the reference part of the Hamiltonian, $H_0$, is accounted for in full and treated analytically whereas only the short-range scattering potential is approximated in a well-defined and controllable manner.

One-dimensional nonrelativistic steady flux scattering could be represented by the solution of the following time-independent Schrödinger equation

$$\left[-\frac{\hbar^2}{2m}\frac{d^2}{dx^2}+V(x)\right]\psi(x,E)=E\,\psi(x,E), \qquad x\in\mathbb{R}, \tag{1.2}$$

where $V(x)=V_0(x)+\tilde{V}(x)$. The "reference potential" $V_0(x)$ is part of the reference Hamiltonian, $H_0=-\frac{\hbar^2}{2m}\frac{d^2}{dx^2}+V_0(x)$, that admits an exact analytic solution. However, the remaining part of the scattering potential, $\tilde{V}(x)$, is assumed to be nonsingular having a finite range ($X$ is the cutoff or range of the potential)

$$\tilde{V}(x)\to 0,\ |x|>X, \tag{1.3}$$

but not required to be analytic. The wavefunction boundary conditions are:

$$\lim_{x\to-\infty}\psi(x,E)=e^{ikx}+Re^{-ikx}, \qquad (1.4)$$

$$\lim_{x\to+\infty}\psi(x,E)=Te^{ikx}, \qquad (1.5)$$

where $k=\sqrt{2mE/\hbar^2}$ ($E$ is the incident beam energy, which is positive and continuous). The incident flux from left is normalized to unity.

We start by solving the reference problem, where $\tilde{V}=0$. Asymptotically, this solution will play the role of the carrier of scattering information in the presence of the potential. If we let $\psi_0(x)$ stand for the reference wavefunction, then

$$\left[-\frac{\hbar^2}{2m}\frac{d^2}{dx^2}+V_0(x)\right]\psi_0(x)=E\,\psi_0(x),\ x\in\mathbb{R}. \qquad (1.6)$$

In this work, we consider the special case where $V_0(x)=0$. Consequently, the two independent real solutions of the reference wave equation (1.6) are $S^+(x)=A\cos(kx)$ which is even and obeying the boundary condition $\left(dS^+/dx\right)_{x=0}=0$ and $S^-(x)=B\sin(kx)$ which is odd and obeying the boundary condition $S^-(x)_{x=0}=0$, where $A$ and $B$ are real constants. From now on, we take $\hbar=m=1$ in all of our equations and numerical computations.

## II. Solution of the J-matrix reference problem

In the J-matrix approach one usually starts by solving the reference problem defined by

$$(H_0-E)|\psi_0\rangle=0. \qquad (2.1)$$

Since this differential equation is of a second order, then one expects two independent solutions. Asymptotically (far away from the scattering center), they are sinusoidal since they represent free particles. The scattering information will be stored in the phase difference between these two sinusoidal waves. It happens that in 3D one of the solutions of the reference problem is regular at the origin while the other is irregular. In the J-matrix formalism, the former solution is referred to as the "sine-like" solution and is expanded in a suitable $L^2$ basis set, $|\phi_n\rangle$, that ensures a tridiagonal representation of the wave operator (2.1). Hence, writing the sine-like solution of (2.1) as $|S\rangle=\sum_n s_n|\phi_n\rangle$ results in the following three-term recursion relation for the expansion coefficients $s_n$

$$J_{n,n}s_n+J_{n,n+1}s_{n+1}+J_{n,n-1}s_{n-1}=0, \qquad (2.2)$$

where $n=0,1,2,\ldots$, $J_{n,m}=\langle\phi_n|(H_0-E)|\phi_m\rangle$, and $J_{0,-1}\equiv 0$. Due to the completeness of the basis, we can construct the other independent *asymptotic* solution of the reference problem, which is complementary to the sine-like solution. We write it as $|C\rangle=\sum_n c_n|\phi_n\rangle$ and refer to it as the "cosine-like" solution. The new expansion coefficients, $c_n$, obey the same recursion relation as $s_n$ except for the initial condition ($n$ = 0) in (2.2). This requirement ensure that $s_n$ and $c_n$ are independent solutions of the three term recursion

relation[‡] and makes the cosine-like solution, $\left|C\right\rangle$, well behaved at the origin. In fact, $\left|C\right\rangle$ and the irregular solution (in 3D) of the reference problem become asymptotically identical [6]. The scattering cross section and phase shift (phase difference between the asymptotic sinusoidal waves $\left|S\right\rangle$ and $\left|C\right\rangle$) will depend on the kinematics and the scattering potential.

In the case of the 1D J-matrix and due to the even parity of the reference Hamiltonian, we will find that the solution space of the reference problem splits into even and odd "J-matrix channels". Thus, we end up with two complementary solutions, one for each channel. The regular or sine-like solutions are defined by

$$\left(H_0 - E\right)\left|S^{\pm}\right\rangle = 0 \tag{2.3}$$

where $\left|S^{\pm}\right\rangle = \sum_n s_n^{\pm}\left|\phi_n^{\pm}\right\rangle$ and the expansion coefficients $s_n^{\pm}$ obey the homogeneous recursion relation (2.2). The cosine-like solutions, $\left|C^{\pm}\right\rangle = \sum_n c_n^{\pm}\left|\phi_n^{\pm}\right\rangle$, will be constructed along the same line as in the 3D case. That is, its expansion coefficients $c_n^{\pm}(E)$ obey the same differential equation in *energy* as do $s_n^{\pm}(E)$. They also satisfy the same recursion relation but with different initial conditions.

Again, the solution of the J-matrix reference problem means writing the analytic reference solution in terms of the elements of a complete discrete basis set in which the matrix representation of the reference wave operator is tridiagonal [3]. That is, the matrix elements of the reference wave operator, $J = H_0 - E$, in the properly chosen basis must satisfy $J_{nm} = 0$ for all indices $\left|n-m\right| \geq 2$. Of course, the basis should be compatible with the domain of $H_0$ and satisfy the boundary conditions. Now, the J-matrix basis that is compatible with our 1D problem and with the solution space for $S^+(x)$ has the following even elements [7]:

$$\phi_n^+(x) = a_n^+ e^{-\lambda^2 x^2/2} H_{2n}(\lambda x), \qquad x \in \mathbb{R}, \tag{2.4}$$

where $H_m(y)$ is the Hermite polynomial of order *m* and $\lambda$ is a real length scale parameter. The normalization constant is conveniently chosen so as to ensure orthonormalization [4] of $\phi_n^+(x)$ giving $a_n^+ = \frac{\sqrt{\lambda/\sqrt{\pi}}}{2^n\sqrt{\Gamma(2n+1)}}$. These functions, $\phi_n^+(x)$, are even in *x* like $S^+(x)$ and satisfy the same condition at the origin; namely, $\left(dS^+/dx\right)_{x=0} = 0$. Using the differential equation, recursion relation, and orthogonality property of the Hermite polynomials [4], we obtain the following

$$\begin{aligned}\left\langle\phi_n^+\right|\left(H_0 - E\right)\left|\phi_m^+\right\rangle \equiv J_{nm}^+(E) = &\frac{\lambda^2}{2}\left[\left(2n+\tfrac{1}{2}\right) - \mu^2\right]\delta_{n,m} \\ &-\frac{\lambda^2}{2}\left[\sqrt{n\left(n-\tfrac{1}{2}\right)}\,\delta_{n,m+1} + \sqrt{(n+1)\left(n+\tfrac{1}{2}\right)}\,\delta_{n,m-1}\right]\end{aligned} \tag{2.5}$$

‡ That is, the Wronskian-like, $w(E) = s_n c_{n-1} - c_n s_{n-1}$, does not vanish and is independent of *n*.

where $\mu = k/\lambda$. The integration measure is $\langle f | g \rangle = \int_{-\infty}^{+\infty} f(x) g(x)\, dx$. We Multiply both sides of the expansion $S^{+}(x) = A\cos(kx) = \sum_{n=0}^{\infty} s_n^{+} \phi_n^{+}(x)$ by $e^{-x^2/2} H_{2m}(x)$ and integrate. Using the orthogonality relation of the Hermite polynomials and the integral formula [8]

$$\int_{-\infty}^{+\infty} e^{-x^2/2} \cos(\mu x) H_{2n}(x) dx = (-1)^n \sqrt{2\pi}\, e^{-\mu^2/2} H_{2n}(\mu), \tag{2.6}$$

we obtain

$$s_n^{+}(E) = \frac{(-1)^n \pi^{1/4} \sqrt{2} A}{2^n \sqrt{\lambda \Gamma(2n+1)}} e^{-\mu^2/2} H_{2n}(\mu). \tag{2.7}$$

Therefore, the reference wave equation $\left(H_0 - E\right)\left|S^{+}\right\rangle = 0$ becomes equivalent to the matrix equation $\sum_m J_{nm}^{+} s_m^{+} = 0$ $, n = 1,2,3...$, where $J_{nm}^{+}$ are the elements of the tridiagonal matrix (2.5). This results in the following three-term recursion relation for the expansion coefficients of the reference wavefunction $S^{+}(x)$

$$\mu^2 s_n^{+} = \left(2n + \tfrac{1}{2}\right) s_n^{+} - \sqrt{n\left(n - \tfrac{1}{2}\right)} s_{n-1}^{+} - \sqrt{(n+1)\left(n + \tfrac{1}{2}\right)} s_{n+1}^{+},\ n = 1,2,3,..., \tag{2.8a}$$

$$\mu^2 s_0^{+} = \tfrac{1}{2} s_0^{+} - \sqrt{\tfrac{1}{2}} s_1^{+}. \tag{2.8b}$$

Now, we turn to $S^{-}(x)$. The J-matrix basis, which is suitable for this solution, has the following odd elements

$$\phi_n^{-}(x) = a_n^{-} e^{-\lambda^2 x^2/2} H_{2n+1}(\lambda x),\quad x \in \mathbb{R}, \tag{2.9}$$

where $a_n^{-} = \frac{\sqrt{\lambda/\sqrt{\pi}}}{2^n \sqrt{2\Gamma(2n+2)}}$. These functions, $\phi_n^{-}(x)$, are odd in *x* like $S^{-}(x)$ and satisfy the same condition at the origin, $S^{-}(0) = 0$. One can easily verify that the two basis elements (2.4) and (2.9) are orthogonal to each other (i.e., $\left\langle \phi_n^{+} \middle| \phi_m^{-} \right\rangle = 0$) and that each form an orthonormal set (i.e., $\left\langle \phi_n^{\pm} \middle| \phi_m^{\pm} \right\rangle = \delta_{nm}$). Moreover, they are related as: $\phi_n^{-}(x) = \phi_{n+1/2}^{+}(x)$. Again, using the properties of the Hermite polynomials, we obtain:

$$\begin{aligned} \left\langle \phi_n^{-} \middle| (H_0 - E) \middle| \phi_m^{-} \right\rangle \equiv J_{nm}^{-}(E) &= \frac{\lambda^2}{2}\left[\left(2n + \tfrac{3}{2}\right) - \mu^2\right]\delta_{n,m} \\ &- \frac{\lambda^2}{2}\left[\sqrt{n\left(n+\tfrac{1}{2}\right)}\,\delta_{n,m+1} + \sqrt{(n+1)\left(n+\tfrac{3}{2}\right)}\,\delta_{n,m-1}\right] \end{aligned} \tag{2.10}$$

Moreover, using the orthogonality property of the Hermite polynomials and the integral formula [8],

$$\int_{-\infty}^{+\infty} e^{-x^2/2} \sin(\mu x) H_{2n+1}(x) dx = (-1)^n \sqrt{2\pi}\, e^{-\mu^2/2} H_{2n+1}(\mu), \tag{2.11}$$

in the expansion $S^{-}(x) = B\sin(kx) = \sum_{n=0}^{\infty} s_n^{-} \phi_n^{-}(x)$, we obtain

$$s_n^{-}(E) = \frac{(-1)^n \pi^{1/4} B}{2^n \sqrt{\lambda \Gamma(2n+2)}} e^{-\mu^2/2} H_{2n+1}(\mu). \tag{2.12}$$

These satisfy the following recursion and its initial relation:

$$\mu^2 s_n^{-} = \left(2n + \tfrac{3}{2}\right) s_n^{-} - \sqrt{n\left(n + \tfrac{1}{2}\right)} s_{n-1}^{-} - \sqrt{(n+1)\left(n + \tfrac{3}{2}\right)} s_{n+1}^{-},\ n = 1,2,3,..., \tag{2.13a}$$

$$\mu^2 s_0^{-} = \tfrac{3}{2} s_0^{-} + \sqrt{\tfrac{3}{2}} s_1^{-}. \tag{2.13b}$$

Therefore, the two coefficients $s_n^{\pm}(E)$ obey two different recursion relations. They are related as follows

$$s_n^-(E) = -\mathrm{i}\tfrac{B}{A}\, s_{n+\frac{1}{2}}^+(E)\,. \qquad (2.14)$$

Moreover, using the differential equation of the Hermite polynomials one can easily verify that $s_n^\pm(E)$ satisfy the following second order differential equation in the energy variable

$$\left[\tfrac{d^2}{d\mu^2} - \mu^2 + 2(2n+1) \mp 1\right] s_n^\pm(\mu) = 0\,. \qquad (2.15)$$

where $\mu = k/\lambda = \frac{1}{\lambda}\sqrt{2E}$ . Additionally, we can in fact determine all expansion coefficients $\left\{s_n^\pm\right\}_{n=0}^\infty$ using only $s_0^\pm(E)$ as seed in the three-term recursion relations (2.8) or (2.13). These seeds are

$$s_0^+(E) = \sqrt{\tfrac{2}{\lambda}}\, A\pi^{1/4} e^{-\mu^2/2}\,,\quad s_0^-(E) = \tfrac{2}{\sqrt{\lambda}}\pi^{1/4} B\,\mu\, e^{-\mu^2/2}\,. \qquad (2.16)$$

Now, we are in a position to search for the J-matrix “complementary” solutions to $S^\pm(x)$. Asymptotically, these solutions, which we denote as $C^\pm(x)$, must be sinusoidal with a phase shift equal to $\pi/2$ as compared to the phases of $S^\pm(x)$. As stated above, we refer to these complementary solutions as the “cosine-like” solutions. They are expanded in the same basis elements $\left\{\phi_n^\pm(x)\right\}$ but with new independent coefficients $c_n^\pm(E)$. These coefficients are required to satisfy the following conditions:

(1) Be an independent solution of the 2nd order energy differential equation (2.15),

(2) Satisfy the same corresponding recursion relation as does $s_n^\pm(E)$ except for the initial relations (2.8b) and (2.13b), and

(3) Make the complementary wavefunctions $C^\pm(x) = \sum_n c_n^\pm(E)\phi_n^\pm(x)$ asymptotically sinusoidal and identical to $S^\pm(x)$ but with a phase difference that depends on the scattering potential parameters and the energy. In the absence of $\tilde{V}$, this phase difference is $\frac{\pi}{2}$.

The first condition is satisfied by the following independent solutions of Eq. (2.15):

$$c_n^+(E) = \frac{A}{\sqrt{\lambda}} D_n^+\, \mu\, e^{-\mu^2/2}\, {}_1F_1\left(-n+\tfrac{1}{2};\tfrac{3}{2};\mu^2\right), \qquad (2.17a)$$

$$c_n^-(E) = \frac{B}{\sqrt{\lambda}} D_n^-\, e^{-\mu^2/2}\, {}_1F_1\left(-n-\tfrac{1}{2};\tfrac{1}{2};\mu^2\right), \qquad (2.17b)$$

where ${}_1F_1(a;c;z)$ is the confluent hypergeometric function [4] and $D_n^\pm$ are normalization constants, which are determined by the second condition as

$$D_n^+ = d^+\sqrt{\Gamma(n+1)\big/\Gamma\left(n+\tfrac{1}{2}\right)}\,, \qquad (2.18a)$$

$$D_n^- = d^-\sqrt{\Gamma(n+1)\big/\Gamma\left(n+\tfrac{3}{2}\right)}\,, \qquad (2.18b)$$

where $d^\pm$ are constants; independent of $\mu$ and $n$. They are determined by the third condition as $d^+ = 2\sqrt{2}$ and $d^- = \sqrt{2}$. This could also be verified analytically and/or numerically. The initial recursion relations satisfied by $c_n^\pm(E)$ are

$$J_{00}^+ c_0^+ + J_{01}^+ c_1^+ = -2^{-\frac{1}{2}}\pi^{-\frac{1}{4}}\lambda^{\frac{3}{2}} A\,\mu\, e^{\mu^2/2}\,, \qquad (2.19a)$$

$$J_{00}^- c_0^- + J_{01}^- c_1^- = \tfrac{1}{2}\pi^{-\frac{1}{4}}\lambda^{\frac{3}{2}} B\, e^{\mu^2/2}\,. \qquad (2.19b)$$

Thus, one needs only $c_0^{\pm}(E)$ as a seed in the three-term recursion relation to obtain all $c_n^{\pm}(E)$ recursively. These seeds are

$$c_0^{+}(E) = 2\sqrt{\tfrac{2}{\lambda}}\,\pi^{-\frac{1}{4}} A\mu e^{-\mu^2/2}\,{}_1F_1\left(\tfrac{1}{2};\tfrac{3}{2};\mu^2\right), \tag{2.20a}$$

$$c_0^{-}(E) = \tfrac{2}{\sqrt{\lambda}}\,\pi^{-\frac{1}{4}} B e^{-\mu^2/2}\,{}_1F_1\left(-\tfrac{1}{2};\tfrac{1}{2};\mu^2\right). \tag{2.20b}$$

The cosine-like functions $C^{\pm}(x)$, which are complementary to the sine-like $S^{\pm}(x)$, have very interesting and intriguing properties and were studied in details by one of the authors [9]. They are trigonometric *only* asymptotically and obey the following asymptotic boundary conditions

$$\lim_{x\to\pm\infty} \tfrac{1}{A} C^{+}(x) = \pm\sin(kx) = \pm\tfrac{1}{B} S^{-}(x), \tag{2.21a}$$

$$\lim_{x\to\pm\infty} \tfrac{1}{B} C^{-}(x) = \pm\cos(kx) = \pm\tfrac{1}{A} S^{+}(x). \tag{2.21b}$$

These properties will proof very valuable in the construction of the asymptotic solution such that it satisfies the boundary conditions (1.4) and (1.5). If we eliminate the lowest terms and construct the following truncated series

$$S_N^{\pm}(x) = \sum_{n=N}^{\infty} s_n^{\pm}(E)\phi_n^{\pm}(x), \tag{2.22a}$$

$$C_N^{\pm}(x) = \sum_{n=N}^{\infty} c_n^{\pm}(E)\phi_n^{\pm}(x), \tag{2.22b}$$

for some large enough integer *N*, then we obtain vanishingly small values for these functions within a finite region symmetric about the origin. The size of this region increases with the number of eliminated terms, *N* [9]. In fact, there is a perfect parallel between the limit as $|x|\to\infty$ of these functions and the limit as $N\to\infty$. Combining (2.21) and (2.22) in a judicious manner so as to obtain sine and cosine functions confined to either end of the real line, we obtain

$$\lim_{N\to\infty}\left[\tfrac{1}{A}S_N^{+}(x) \pm \tfrac{1}{B}C_N^{-}(x)\right] = \lim_{|x|\to\infty}\left[\tfrac{1}{A}S^{+}(x) \pm \tfrac{1}{B}C^{-}(x)\right] = \begin{cases} 2\cos(kx) & ,x\to\pm\infty \\ 0 & ,x\to\mp\infty \end{cases} \tag{2.23a}$$

$$\lim_{N\to\infty}\left[\tfrac{1}{B}S_N^{-}(x) \pm \tfrac{1}{A}C_N^{+}(x)\right] = \lim_{|x|\to\infty}\left[\tfrac{1}{B}S^{-}(x) \pm \tfrac{1}{A}C^{+}(x)\right] = \begin{cases} 2\sin(kx) & ,x\to\pm\infty \\ 0 & ,x\to\mp\infty \end{cases} \tag{2.23b}$$

Figure 1a and 1b show these partial sums for the top and bottom signs, respectively. The number of truncated terms are $N = 0, 10, 30$.

In the following section, the J-matrix scheme will be used to augment the kinematics (solutions of the reference problem) obtained above by the dynamics coming from the contribution of the potential $\tilde{V}(x)$. The asymptotics obtained above will then be used to match the boundary conditions and identify the corresponding scattering amplitudes *R*(*E*) and *T*(*E*).

## III. The J-matrix calculation of the reflection and transmission amplitudes

We assume that the scattering potential $\tilde{V}(x)$ has a short range such that it is accurately represented by its matrix elements in a finite subset of the basis. This basis must be complete and compatible with the 1D problem so that it carries a faithful representation of the total wavefunction $\psi(x,E)$ that solves Eq. (1.2). That is, we can write $\psi(x,E)$ as an infinite convergent series in this complete basis set

$$\psi(x,E)=\sum_{m=-\infty}^{+\infty} a_m(E)\xi_m(x)\,; \quad \sum_{m=-\infty}^{+\infty} |\xi_m\rangle\langle\xi_m| = 1 \tag{3.1}$$

The relation between $\{\xi_m\}_{m=-\infty}^{\infty}$ and $\{\phi_n^{\pm}\}_{n=0}^{\infty}$ will be given shortly below. All elements of the basis (odd and even) are necessary even if the scattering potential function has a definite parity. For example, if $\tilde{V}(x)$ is an even function then the potential matrix elements $\langle\phi_n^{\pm}|\tilde{V}|\phi_m^{\pm}\rangle$ are non-vanishing. On the other hand, if $\tilde{V}(x)$ is odd then the non-zero matrix elements are $\langle\phi_n^{\pm}|\tilde{V}|\phi_m^{\mp}\rangle$.

Due to the finite range of the scattering potential, we can split the configuration space as follows:

$$H(x)=\begin{cases} H_0(x) & ,x\le -X \\ H_0(x)+\tilde{V}(x) & ,-X<x<X \\ H_0(x) & ,x\ge X \end{cases}, \quad \psi(x)=\begin{cases} \Psi_0^-(x) & ,x\le -X \\ \Psi(x) & ,-X<x<X \\ \Psi_0^+(x) & ,x\ge X \end{cases} \tag{3.2}$$

where $X$ is the cutoff or range of the potential and $\Psi_0^{\pm}(x)$ are the reference wavefunctions with the added scattering information. The equivalent J-matrix formalism in function space goes as follows. We break down the whole function space into "left", "right", and "middle" subspaces. The left and right subspaces are analogous to the "outer" subspace in the original 3D J-matrix formulation, whereas the middle subspace parallels the "inner" scattering subspace in 3D [3]. Therefore, in analogy to the separation of configuration space in (3.2), we can write the representation of the total Hamiltonian and wavefunction in function space as follows [10]

$$\langle\xi_n|H|\xi_m\rangle=\begin{cases} \langle\xi_n|H_0|\xi_m\rangle & ;n,m<-N \\ \langle\xi_n|\left(H_0+\tilde{V}\right)|\xi_m\rangle & ;-N\le n,m<N \\ \langle\xi_n|H_0|\xi_m\rangle & ;n,m\ge N \end{cases} \tag{3.3}$$

$$\psi(x,E)=\sum\nolimits_{m=-\infty}^{-N-1} b_{-m-1}^-(E)\xi_m(x)+\sum_{m=-N}^{N-1} a_m(E)\xi_m(x)+\sum\nolimits_{m=N}^{+\infty} b_m^+(E)\xi_m(x). \tag{3.4}$$

The apparent asymmetry between the subscripts on the expansion coefficients of the first and last term is put by hand to ensure that once the first term is transformed to a positive summation from $N$ to $\infty$ the expansion coefficients will be denoted by $b_n^-$ for ease of identification with the suitable asymptotics later on. In the above summation the integer $N$ should be large enough such that an accurate and faithful representation is obtained:

(1) For the scattering potential $\tilde{V}(x)$ in the $2N$ dimensional "middle" subspace $\{\xi_m\}_{m=-N}^{N-1}$, and

(2) For the asymptotic sinusoidal limit of the wavefunction $\psi(x,E)$ in the "left" subspace, $\{\xi_m\}_{m=-N-1}^{-\infty}$, as $x\to-\infty$ and in the "right" subspace, $\{\xi_m\}_{m=N}^{\infty}$, as $x\to+\infty$.

Now, in the left subspace, the asymptotic form of the total wavefunction $\psi(x,E)$ is $e^{ikx}+Re^{-ikx}$. On the other hand, the asymptotic form of $\psi(x,E)$ in the right subspace is $Te^{ikx}$. Using the results obtained at the end of the previous section (2.23a-b) on the asymptotic solutions of the reference problem, we can write

$$\lim_{N\to\infty}\left\{\tfrac{1}{2}\left[\tfrac{1}{A}S_N^+(x)\pm\tfrac{1}{B}C_N^-(x)\right]+\tfrac{\mathrm{i}}{2}\left[\tfrac{1}{B}S_N^-(x)\pm\tfrac{1}{A}C_N^+(x)\right]\right\}=\begin{cases}e^{\mathrm{i}kx} & ,x\to\pm\infty\\ 0 & ,x\to\mp\infty\end{cases} \tag{3.5a}$$

$$\lim_{N\to\infty}\left\{\tfrac{1}{2}\left[\tfrac{1}{A}S_N^+(x)\pm\tfrac{1}{B}C_N^-(x)\right]-\tfrac{\mathrm{i}}{2}\left[\tfrac{1}{B}S_N^-(x)\pm\tfrac{1}{A}C_N^+(x)\right]\right\}=\begin{cases}e^{-\mathrm{i}kx} & ,x\to\pm\infty\\ 0 & ,x\to\mp\infty\end{cases} \tag{3.5b}$$

Based on these observations we can write the asymptotic solutions of the total wave function as follows

$$\lim_{x\to+\infty}\psi(x,E)=\Psi_0^+=Te^{ikx}=T\sum_{n=N}^{\infty}\left[\tfrac{1}{2A}\left(s_n^++\mathrm{i}c_n^+\right)\phi_n^++\tfrac{1}{2B}\left(c_n^-+\mathrm{i}s_n^-\right)\phi_n^-\right], \tag{3.6}$$

$$\begin{aligned}\lim_{x\to-\infty}\psi(x,E)=\Psi_0^-=e^{ikx}+Re^{-ikx}&=\sum_{n=N}^{\infty}\left[\tfrac{1}{2A}\left(s_n^+-\mathrm{i}c_n^+\right)\phi_n^+-\tfrac{1}{2B}\left(c_n^--\mathrm{i}s_n^-\right)\phi_n^-\right]\\ &+R\sum_{n=N}^{\infty}\left[\tfrac{1}{2A}\left(s_n^++\mathrm{i}c_n^+\right)\phi_n^+-\tfrac{1}{2B}\left(c_n^-+\mathrm{i}s_n^-\right)\phi_n^-\right]\end{aligned} \tag{3.7}$$

Comparing these two asymptotic solutions represented as summations with the corresponding sum in Eq. (3.4) we deduce that

$$\xi_m(x)=\begin{cases}\phi_m^+(x) & ,\quad m\ge 0\\ \phi_{-m-1}^-(x) & ,\quad m<0\end{cases} \tag{3.8}$$

$$b_n^+(E)=W_+(E)f_n^+(E)+f_n^-(E), \tag{3.9a}$$

$$b_n^-(E)=W_-(E)g_n^+(E)-g_n^-(E), \tag{3.9b}$$

where $W_\pm(E)=T\pm R$, $f_n^\pm(E)=\frac{1}{2A}\left(s_n^+\pm\mathrm{i}c_n^+\right)$, and $g_n^\pm(E)=\frac{1}{2B}\left(c_n^-\pm\mathrm{i}s_n^-\right)$. Consequently, a solution of the problem (i.e., complete specification of the total wave function $\psi$) is obtained if we could determine the $2N+2$ energy-dependent quantities, $\left\{a_m(E)\right\}_{m=-N}^{N-1}$ and $W_\pm(E)$. Now, the action of the wave operator on the total wave function (3.4) reads as follows

$$\begin{aligned}(H-E)|\psi\rangle&=\sum_{m=-N}^{N-1}a_m(H-E)|\xi_m\rangle\\ &+\sum\nolimits_{m=N}^{+\infty}b_m^+(H_0-E)|\xi_m\rangle+\sum\nolimits_{m=-\infty}^{-N-1}b_{-m-1}^-(H_0-E)|\xi_m\rangle=0\end{aligned} \tag{3.10}$$

Projecting on the left by $\langle\xi_n|$, we obtain

$$\langle\xi_n|(H-E)|\psi\rangle=\sum_{m=-N}^{N-1}\left(J_{nm}+\tilde{V}_{nm}\right)a_m+\sum\nolimits_{m=N}^{+\infty}J_{nm}^+b_m^++\sum\nolimits_{m=-\infty}^{-N-1}J_{-n-1,-m-1}^-b_{-m-1}^-=0, \tag{3.11}$$

where again $J=H_0-E$ and we have noted that for $n,m\ge N$ we obtain $\langle\xi_n|J|\xi_m\rangle=\langle\phi_n^+|J|\phi_m^+\rangle=J_{nm}^+$ whereas for $n,m\le -N-1$ the result is $\langle\xi_n|J|\xi_m\rangle=\langle\phi_{-n-1}^-|J|\phi_{-m-1}^-\rangle=J_{-n-1,-m-1}^-$. Moreover, for positive/negative index $n$ the third/second sum in Eq. (3.11) disappears since the wave operator $J$ is even. The matrix elements of a general function of (or operator in) $x$, $F(x)$, in the "middle" subspace $\left\{\xi_m\right\}_{m=-N}^{N-1}$ is evaluated as follows

$$F_{nm}=\langle\xi_n|F|\xi_m\rangle=\begin{cases}F^{++}_{n,m}=\langle\phi^+_n|F^+|\phi^+_m\rangle & ;0\le n,m\le N-1\\ F^{--}_{-n-1,-m-1}=\langle\phi^-_{-n-1}|F^+|\phi^-_{-m-1}\rangle & ;-1\ge n,m\ge -N\\ F^{+-}_{n,-m-1}=\langle\phi^+_n|F^-|\phi^-_{-m-1}\rangle & ;0\le n\le N-1,-1\ge m\ge -N\\ F^{-+}_{-n-1,m}=\langle\phi^-_{-n-1}|F^-|\phi^+_m\rangle & ;-1\ge n\ge -N,0\le m\le N-1\end{cases}\tag{3.12}$$

where $F^{\pm}(x)=\frac{1}{2}\left[F(x)\pm F(-x)\right]$ is the even/odd part of $F(x)$. Therefore, since the reference wave operator is even, then we obtain its matrix representation in the middle subspace as follows

$$J_{n,m}=\begin{cases}J^+_{n,m} & ;n,m\ge 0\\ J^-_{-n-1,-m-1} & ;n,m\le -1\\ 0 & ;\text{otherwise}\end{cases}\tag{3.13}$$

Figure 2 is a schematic diagram for the infinite tridiagonal matrix representation of the reference wave operator in the whole space. One should note that $J_{0,-1}=J_{-1,0}=0$. Moreover, for a general scattering potential function $\tilde{V}(x)$ with no well-defined parity the representation of the total Hamiltonian is shown schematically in Fig. 3. The finite $2N\times 2N$ symmetric matrix block on the diagonal is the representation of the total Hamiltonian in the "middle" subspace spanned by $\{\xi_m\}_{m=-N}^{N-1}$. Fig. 4 shows that this Hamiltonian matrix block is made up, in fact, of four sub-blocks. The two $N\times N$ diagonal sub-blocks represent the even part of the finite Hamiltonian matrix, whereas, the two $N\times N$ off-diagonal sub-blocks represent the odd part. If the scattering potential $\tilde{V}(x)$ is an even function of $x$, then the two off-diagonal sub-blocks become zero and the whole problem is completely reducible into two semi-infinite problems. However, if $\tilde{V}(x)$ is an odd function of $x$, then in this case, the two diagonal sub-blocks will be represented by the tridiagonal structure of the reference wave operator $J$ shown in Fig. 2.

We should also note here that, in general, the cutoff parameter of the scattering potential might not be symmetric. That is, $\tilde{V}(x)=0$ for $\pm x\ge X^{\pm}$, where $X^+\ne X^-$. In such a case, the configuration space representation (3.3) of the total Hamiltonian becomes

$$H(x)=\begin{cases}H_0(x) & ,x\le -X^-\\ H_0(x)+\tilde{V}(x) & ,-X^-<x<X^+\\ H_0(x) & ,x\ge X^+\end{cases}\tag{3.14}$$

The J-matrix equivalent formulation for this case gives the following matrix representation of the total Hamiltonian

$$\langle\xi_n|H|\xi_m\rangle=\begin{cases}\langle\xi_n|H_0|\xi_m\rangle & ;n,m<-N^-\\ \langle\xi_n|\left(H_0+\tilde{V}\right)|\xi_m\rangle & ;-N^-\le n,m<N^+\\ \langle\xi_n|H_0|\xi_m\rangle & ;n,m\ge N^+\end{cases}\tag{3.15}$$

where $N^+\ne N^-$. Moreover, we are also at liberty to choose two different scale parameters $\lambda^{\pm}$ for the even and odd channels spanned by $\{\phi^{\pm}_n\}$, respectively. Figure 5 is a reproduction of Fig. 4 for such non-symmetric case. However, we limit our present study to the case where $N^+=N^-=N$ and $\lambda^+=\lambda^-=\lambda$. Next, we investigate the solution of the wave equation $\langle \xi_n | (H-E) | \psi \rangle = 0$ with the help of its matrix representation in Eq. (3.11) for all *n*. The goal is to calculate the two energy-dependent amplitudes *T*(*E*), and *R*(*E*) (or, equivalently, $W_\pm$).

It is easy to see that the coefficients $b_n^\pm(E)$ defined in Eq. (3.9) satisfy the following three-term recursion relation for $n \geq 1$

$$J_{n,n-1}^{+} b_{n-1}^{+} + J_{n,n}^{+} b_{n}^{+} + J_{n,n+1}^{+} b_{n+1}^{+} = 0, \tag{3.16a}$$

$$J_{n,n-1}^{-} b_{n-1}^{-} + J_{n,n}^{-} b_{n}^{-} + J_{n,n+1}^{-} b_{n+1}^{-} = 0. \tag{3.16b}$$

Therefore, the wave equation $\langle \xi_n | (H-E) | \psi \rangle = 0$ is automatically satisfied for $n \geq N+1$ and $n \leq -N-2$. On the other hand, for $n = N$ and $n = -N-1$, the wave equation gives

$$a_{N-1}(E) = b_{N-1}^{+}(E), \text{ and } a_{-N}(E) = b_{N-1}^{-}(E), \tag{3.17}$$

respectively. Whereas, for $n = N-1$ and $n = -N$, we obtain the following two equations

$$\sum_{m=-N}^{N-1} \left( J_{N-1,m} + \tilde{V}_{N-1,m} \right) a_m = -J_{N-1,N}^{+} b_N^{+}, \tag{3.18a}$$

$$\sum_{m=-N}^{N-1} \left( J_{-N,m} + \tilde{V}_{-N,m} \right) a_m = -J_{N-1,N}^{-} b_N^{-}. \tag{3.18b}$$

However, for $N-2 \geq n \geq -N+1$, the matrix wave equation simplifies to

$$\sum_{m=-N}^{N-1} \left( J_{nm} + \tilde{V}_{nm} \right) a_m = 0. \tag{3.19}$$

Putting together these results obtained in Eqs. (3.17) to (3.19) and using Eqs. (3.12) & (3.13), we arrive at an algebraic representation of the wave equation in the following matrix form

$$\left[ \begin{pmatrix} J^{-} & 0 \\ 0 & J^{+} \end{pmatrix} + \begin{pmatrix} \tilde{V}_{\text{even}}^{--} & \tilde{V}_{\text{odd}}^{-+} \\ \tilde{V}_{\text{odd}}^{+-} & \tilde{V}_{\text{even}}^{++} \end{pmatrix} \right] \bullet \begin{bmatrix} b_{N-1}^{-} \\ a_{-N+1} \\ \vdots \\ a_{N-2} \\ b_{N-1}^{+} \end{bmatrix} = \begin{bmatrix} -J_{N-1,N}^{-} b_N^{-} \\ 0 \\ \vdots \\ 0 \\ -J_{N-1,N}^{+} b_N^{+} \end{bmatrix} \tag{3.20}$$

More details on this matrix equation are shown in Fig. 6. The results are 2*N* linear equations with 2*N* unknowns, $\{a_m(E)\}_{m=-N+1}^{N-2}$, *T*(*E*), and *R*(*E*). We define the finite 2*N*×2*N* matrix Green's function as

$$[G(E)]_{nm} = \left[ (H-E)^{-1} \right]_{nm} = \left[ \left( H_0 + \tilde{V} - E \right)^{-1} \right]_{nm} = \left[ \left( J + \tilde{V} \right)^{-1} \right]_{nm}, \tag{3.21}$$

where $\{n, m\} = -N, -N+1, ..., N-2, N-1$. One of the advantages of the J-matrix method is that the diagonalization of the finite total Hamiltonian matrix $\{H_{nm}\}_{n,m=-N}^{N-1}$, which is needed in the evaluation of *G*(*E*), is done only once and for all energies. Moreover, another advantage is our choice of basis that gives the following simple form for the components of *G*(*E*) that enter in the calculation of *R*(*E*) and *T*(*E*) [11]:

$$G_{nm}(E) = \sum_{k=-N}^{N-1} \frac{\Lambda_{nk}\Lambda_{mk}}{\varepsilon_k - E}, \tag{3.22}$$

where $\{\varepsilon_k\}_{k=-N}^{N-1}$ are the 2*N* energy eigenvalues of the finite total Hamiltonian matrix *H* and $\{\Lambda_{nk}\}_{n,k=-N}^{N-1}$ are the corresponding normalized eigenvectors. Using the matrix Green's function *G*(*E*), we rewrite Eq. (3.20) as

$$\begin{bmatrix} b_{N-1}^- \\ a_{-N+1} \\ \vdots \\ a_{N-2} \\ b_{N-1}^+ \end{bmatrix} = \left( \quad G(E) \quad \right) \bullet \begin{bmatrix} -J_{N-1,N}^- b_N^- \\ 0 \\ \vdots \\ 0 \\ -J_{N-1,N}^+ b_N^+ \end{bmatrix} \tag{3.23}$$

This gives the following two equations relating and containing only the two unknowns *T*(*E*), and *R*(*E*):

$$-b_{N-1}^- = G_{-N,-N} J_{N-1,N}^- b_N^- + G_{-N,N-1} J_{N-1,N}^+ b_N^+ , \tag{3.24a}$$

$$-b_{N-1}^+ = G_{N-1,N-1} J_{N-1,N}^+ b_N^+ + G_{N-1,-N} J_{N-1,N}^- b_N^- . \tag{3.24b}$$

Using the definition of $b_m^{\pm}(E)$ given by Eq. (3.9) and after some manipulations, we obtain

$$\begin{aligned} W_+(E) = & \left[ 1 - \frac{\alpha_N^+ \beta_N^+ \mathcal{J}_+ \mathcal{J}_- \mathcal{G}_{+-} \mathcal{G}_{-+}}{\left(1+\mathcal{G}_{++}\mathcal{J}_+\alpha_N^+\right)\left(1+\mathcal{G}_{--}\mathcal{J}_-\beta_N^+\right)} \right]^{-1} \left[ -\rho_{N-1} \frac{1+\mathcal{G}_{++}\mathcal{J}_+\alpha_N^-}{1+\mathcal{G}_{++}\mathcal{J}_+\alpha_N^+} \right. \\ & \left. + \frac{\alpha_N^+}{\gamma_N^+} \frac{\mathcal{G}_{+-}\mathcal{J}_-}{1+\mathcal{G}_{++}\mathcal{J}_+\alpha_N^+} \left( \sigma_N - \sigma_{N-1} \frac{1+\mathcal{G}_{--}\mathcal{J}_-\beta_N^-}{1+\mathcal{G}_{--}\mathcal{J}_-\beta_N^+} + \rho_N \frac{\gamma_N^+ \beta_N^+ \mathcal{G}_{-+} \mathcal{J}_+}{1+\mathcal{G}_{--}\mathcal{J}_-\beta_N^+} \right) \right] \end{aligned} \tag{3.25a}$$

$$\begin{aligned} W_-(E) = & \left[ 1 - \frac{\alpha_N^+ \beta_N^+ \mathcal{J}_+ \mathcal{J}_- \mathcal{G}_{+-} \mathcal{G}_{-+}}{\left(1+\mathcal{G}_{++}\mathcal{J}_+\alpha_N^+\right)\left(1+\mathcal{G}_{--}\mathcal{J}_-\beta_N^+\right)} \right]^{-1} \left[ \sigma_{N-1} \frac{1+\mathcal{G}_{--}\mathcal{J}_-\beta_N^-}{1+\mathcal{G}_{--}\mathcal{J}_-\beta_N^+} \right. \\ & \left. - \beta_N^+ \gamma_N^+ \frac{\mathcal{G}_{-+}\mathcal{J}_+}{1+\mathcal{G}_{--}\mathcal{J}_-\beta_N^+} \left( \rho_N - \rho_{N-1} \frac{1+\mathcal{G}_{++}\mathcal{J}_+\alpha_N^-}{1+\mathcal{G}_{++}\mathcal{J}_+\alpha_N^+} + \sigma_N \frac{\alpha_N^+ \mathcal{G}_{+-} \mathcal{J}_- / \gamma_N^+}{1+\mathcal{G}_{++}\mathcal{J}_+\alpha_N^+} \right) \right] \end{aligned} \tag{3.25b}$$

where we have defined: $\mathcal{G}_{++} = G_{N-1,N-1}$, $\mathcal{G}_{--} = G_{-N,-N}$, $\mathcal{G}_{+-} = G_{N-1,-N}$, $\mathcal{G}_{-+} = G_{-N,N-1}$, and $\mathcal{J}_\pm = J_{N-1,N}^\pm = -\frac{\lambda^2}{2}\sqrt{N\left(N \mp \frac{1}{2}\right)}$. We have also introduced the following ratios

$$\alpha_n^\pm = f_n^\pm / f_{n-1}^\pm , \; \beta_n^\pm = g_n^\pm / g_{n-1}^\pm , \; \rho_n = f_n^- / f_n^+ , \; \sigma_n = g_n^- / g_n^+ , \; \gamma_n^\pm = f_n^\pm / g_n^\pm . \tag{3.26}$$

All terms in the pair of equations (3.25) that contain $\mathcal{G}_{\pm\mp}$ or $\gamma_n^\pm$ indicate coupling of the odd and even scattering channels. In Appendix A, we show how to calculate the ratios defined in Eq. (3.26) by starting from the seeds $\left\{s_0^\pm, s_1^\pm, c_0^\pm, c_1^\pm\right\}$ and using an efficient, stable, and highly accurate computational scheme. These are written as continued fractions derived from the three-term recursion relations (2.8) and (2.13). The elements of the Green's function matrix are evaluated using formula (3.22). This requires an accurate evaluation of the elements of the 2*N*×2*N* total Hamiltonian matrix $H_0 + \tilde{V}$. It is straightforward to obtain the exact values of the matrix elements of $H_0$ in the basis (3.8) by using Eq. (2.5) and Eq. (2.10) with *E* = 0. Thus, the only remaining quantities are the matrix elements of the short range scattering potential $\tilde{V}$. These could be evaluated to the desired accuracy by using the Gauss quadrature integral approximation as shown in Appendix B. Now, if the scattering potential $\tilde{V}(x)$ is an even function in *x* then $G_{N-1,-N} = G_{-N,N-1} = 0$ and the pair of equations (3.25a) and (3.25b) simplify giving the following transmission and reflection amplitudes

$$T(E) = \tfrac{1}{2}\left(W_+ + W_-\right) = -\frac{1}{2}\left[\rho_{N-1}\frac{1+\mathcal{G}_{++}\mathcal{J}_+\alpha_N^-}{1+\mathcal{G}_{++}\mathcal{J}_+\alpha_N^+} - \sigma_{N-1}\frac{1+\mathcal{G}_{--}\mathcal{J}_-\beta_N^-}{1+\mathcal{G}_{--}\mathcal{J}_-\beta_N^+}\right], \tag{3.27a}$$

$$R(E) = \tfrac{1}{2}\left(W_+ - W_-\right) = -\frac{1}{2}\left[\rho_{N-1}\frac{1+\mathcal{G}_{++}\mathcal{J}_+\alpha_N^-}{1+\mathcal{G}_{++}\mathcal{J}_+\alpha_N^+} + \sigma_{N-1}\frac{1+\mathcal{G}_{--}\mathcal{J}_-\beta_N^-}{1+\mathcal{G}_{--}\mathcal{J}_-\beta_N^+}\right]. \tag{3.27b}$$

It is interesting to note that each term inside the square brackets is a ratio of a complex number and its conjugate. Therefore, these formulas are, in fact, a proof that for any hermitian system with short-range even potential, the transmission and reflection amplitudes can always be written as

$$T = \tfrac{1}{2}\left(e^{2i\theta_+} - e^{2i\theta_-}\right),\; R = \tfrac{1}{2}\left(e^{2i\theta_+} + e^{2i\theta_-}\right). \tag{3.28}$$

where $W_\pm = \pm e^{2i\theta_\pm}$ and $\theta_\pm(E)$ could be interpreted as some scattering phase shift angle, which is a combination of reflection and transmission phase information.

It is important to note that (similar to the standard 3D J-matrix) the physical results obtained will be independent of the value of the length scale parameter $\lambda$ as long as its value is within the plateau of stable computations. Moreover, the size of the plateau of stability increases with *N*, and as $N \to \infty$ the results should be independent of any choice of value for $\lambda$. In the following section, we use the above formulation of the 1D J-matrix method of scattering to calculate the transmission and reflection amplitudes for few examples and compare our results with exact values whenever available.

## IV. Numerical Computations

In this section, we implement the theoretical formulation made above on three examples; two of them are exactly solvable quantum problems. The first case is the so-called modified hyperbolic Pöschl-Teller potential defined by [12]

$$V(x) = -\frac{\eta^2}{2}\frac{\nu(\nu-1)}{\cosh^2 \eta x}, \tag{4.1}$$

which describes an attractive potential for $|\nu| > 1$ and repulsive barrier for $0 < \nu < 1$. The transmission and reflection amplitudes are given by [12]

$$T(E) = \tfrac{1}{2}\left(e^{2i\theta_+} - e^{2i\theta_-}\right),\; R(E) = \tfrac{1}{2}\left(e^{2i\theta_+} + e^{2i\theta_-}\right), \tag{4.2}$$

where

$$\theta_\pm = \arg\left[\Gamma\left(i\tfrac{k}{\eta}\right)e^{-i\frac{k}{\eta}\log 2} \Big/ \Gamma\left(\tfrac{ik}{2\eta} + \tfrac{2\nu-1\pm 1}{4}\right)\Gamma\left(\tfrac{ik}{2\eta} + \tfrac{3\mp 1-2\nu}{4}\right)\right]. \tag{4.3}$$

Moreover, we can also write the simple form

$$|T(E)|^2 = \left(1+p^{-2}\right)^{-1},\; |R(E)|^2 = \left(1+p^2\right)^{-1}, \tag{4.4}$$

where $p = \sinh\left(\pi\tfrac{k}{\eta}\right)\big/\sin(\nu\pi)$. Figure 7 shows a successful comparison of these exact results with those obtained from Eq. (3.27) since the potential (4.1) is even in *x*.

In the second example we consider the classic 1D problem of a square barrier of height $V_0$ and width *L*. To test the general results in Eq. (3.25), we take the barrier location at $0 \le x \le L$ so that it does not have a definite parity. The exact transmission and reflection amplitudes are given by [12]

$$|T(E)|^2 = \left(1+q^2\right)^{-1}, \quad |R(E)|^2 = \left(1+q^{-2}\right)^{-1}, \tag{4.5}$$

where $q = \frac{V_0}{k\hat{k}} \times \begin{cases} \sinh(\hat{k}L) & ,E<V_0 \\ \sin(\hat{k}L) & ,E>V_0 \end{cases}$ and $\hat{k} = \sqrt{2|E-V_0|}$. Figure 8 compares successfully these exact results with those obtained from the general formula in Eq. (3.25).

The third and final example is for a potential that does not have an exact solution. We consider the following

$$V(x) = \begin{cases} V_0 \sin^2\left(\pi x/a\right) & ,|x| \le a \\ 0 & ,\text{otherwise} \end{cases} \tag{4.6}$$

which is a symmetric double barrier potential. Figure 9 shows the transmission and reflection amplitudes for the chosen parameters where it clearly indicates the presence of a sharp resonance at about $E = 0.7V_0$.

These results (and much more extensive list of problems that has been worked out by the authors) demonstrate the efficiency and accuracy of the method making it a viable alternative to the classical treatment of one-dimensional scattering.

## V. Conclusion

In this work, we have extended the standard 3D J-matrix treatment of scattering to the one-dimensional case. The traditional reflection and transmission amplitudes have been expressed in terms of the J-matrix kinematic coefficients and the scattering potential matrix elements in a convenient square integrable basis. The usual thought-off simplicity of the one-dimensional scattering problem as compared to three-dimensional one could be debated as evidenced by the rich structure presented above (e.g., the presence of odd an even coupled channels as opposed a single channel in 3D). However, one of the most important features of the 1D J-matrix method is the fact that it translates all standard concepts, such as the interior and exterior regions, boundary conditions…etc., from the coordinate representation to an algebraic discrete basis representation. These assertions are at the basis of the original formulation of the J-matrix approach. All physical information required to study one-dimensional scattering is carried by the expansion coefficients of the asymptotic wavefunction and the potential matrix elements in the square integrable basis.

The numerical implementation of our theoretical formulation for three test potentials is satisfactory and suggest that the present formulation of the 1D J-matrix can be used as a new alternative to resolve issues related to bound states, resonances and scattering phenomena in one dimensional systems. Moreover, the formulation could be extended to multi-channel as well as relativistic scattering. Work on the latter is under way. Our present success in formulating the J-matrix approach to scattering in 1D give us confidence that its formulation in 2D should be feasible and that our efforts in this direction should culminate to validate the J-matrix method of scattering to all three dimensions.

**Acknowledgements:** Partial support of this work by King Fahd University of Petroleum and Minerals is under project SB-090001 is deeply appreciated. The financial support by the Saudi Center for Theoretical Physics (SCTP) is highly acknowledged.

## Appendix A: Calculating the ratios $\alpha_n^\pm$, $\beta_n^\pm$, $\gamma_n^\pm$, $\rho_n$, and $\sigma_n$ defined in Eq. (3.26)

The coefficients $f_n^\pm(E)$ defined in Eq. (3.9) satisfy the same recursion relation as $s_n^+(E)$ and $c_n^+(E)$. That is,

$$\mu^2 f_n^\pm = \left(2n+\tfrac{1}{2}\right) f_n^\pm - \sqrt{n\left(n-\tfrac{1}{2}\right)}\, f_{n-1}^\pm - \sqrt{(n+1)\left(n+\tfrac{1}{2}\right)}\, f_{n+1}^\pm,\quad n\geq 1 \tag{A.1}$$

Dividing by $f_n^\pm$, we obtain the following

$$\alpha_{n+1}^\pm(\mu) = \frac{2n+\frac{1}{2}-\mu^2}{\sqrt{(n+1)(n+1/2)}} - \frac{1}{\alpha_n^\pm(\mu)}\sqrt{\frac{n(n-1/2)}{(n+1)(n+1/2)}},\qquad n\geq 1 \tag{A.2}$$

Therefore, starting with $\alpha_1^\pm(\mu)$, which depends on $\left\{s_0^+, s_1^+, c_0^+, c_1^+\right\}$, we can compute all higher degree ratios $\left\{\alpha_n^\pm\right\}_{n\geq 2}$. Moreover, Eq. (A.2) results in the following continued fraction for $\alpha_n^\pm$ in terms of $\alpha_1^\pm$ [5]

$$\sqrt{n\left(n-\tfrac{1}{2}\right)}\,\alpha_n^\pm(\mu) = \mu^2-\left(2n-\tfrac{3}{2}\right)-\cfrac{(n-1)(n-3/2)}{\mu^2-\left(2n-\tfrac{7}{2}\right)-\cfrac{(n-2)(n-5/2)}{\mu^2-\left(2n-\tfrac{11}{2}\right)-\dots-\cfrac{15/2}{\mu^2-\frac{9}{2}-\cfrac{3}{\mu^2-\frac{5}{2}-\cfrac{\sqrt{1/2}}{\alpha_1^\pm(\mu)}}}}} \tag{A.3}$$

Similarly, the coefficients $g_n^\pm(E)$ defined by Eq. (3.9) satisfy the following three-term recursion relation

$$\mu^2 g_n^\pm = \left(2n+\tfrac{3}{2}\right) g_n^\pm - \sqrt{n\left(n+\tfrac{1}{2}\right)}\, g_{n-1}^\pm - \sqrt{(n+1)\left(n+\tfrac{3}{2}\right)}\, g_{n+1}^\pm\quad,\quad n\geq 1 \tag{A.4}$$

resulting in the following recursive relation:

$$\beta_{n+1}^\pm(\mu) = \frac{2n+\frac{3}{2}-\mu^2}{\sqrt{(n+1)(n+3/2)}} - \frac{1}{\beta_n^\pm(\mu)}\sqrt{\frac{n(n+1/2)}{(n+1)(n+3/2)}},\qquad n\geq 1 \tag{A.5}$$

Consequently, starting with $\beta_1^\pm$, which depends on $\left\{s_0^-, s_1^-, c_0^-, c_1^-\right\}$, we can determine all $\left\{\beta_n^\pm\right\}_{n\geq 2}$ recursively. Now, the coefficient ratios $\gamma_n^\pm$, $\rho_n(\mu)$, and $\sigma_n(\mu)$ could be rewritten as

$$\gamma_n^\pm(\mu) = \frac{f_n^\pm}{g_n^\pm} = \frac{f_n^\pm}{f_{n-1}^\pm}\frac{f_{n-1}^\pm}{g_{n-1}^\pm}\frac{g_{n-1}^\pm}{g_n^\pm} = \frac{\alpha_n^\pm(\mu)}{\beta_n^\pm(\mu)}\gamma_{n-1}^\pm(\mu), \tag{A.6}$$

$$\rho_n(\mu) = \frac{f_n^-}{f_n^+} = \frac{f_n^-}{f_{n-1}^-}\frac{f_{n-1}^-}{f_{n-1}^+}\frac{f_{n-1}^+}{f_n^+} = \frac{\alpha_n^-(\mu)}{\alpha_n^+(\mu)}\rho_{n-1}(\mu), \tag{A.7}$$

$$\sigma_n(\mu) = \frac{g_n^-}{g_n^+} = \frac{g_n^-}{g_{n-1}^-}\frac{g_{n-1}^-}{g_{n-1}^+}\frac{g_{n-1}^+}{g_n^+} = \frac{\beta_n^-(\mu)}{\beta_n^+(\mu)}\sigma_{n-1}(\mu). \tag{A.8}$$

Therefore, starting with $\gamma_0^\pm$, $\rho_0$ and $\sigma_0$ and using $\{\alpha_n^\pm, \beta_n^\pm\}$ calculated above we obtain all $\{\gamma_n^\pm, \rho_n, \sigma_n\}_{n\geq 1}$. Thus, the sequence of steps for calculating these coefficient ratios goes as follows:

(1) For a given $E$ and $\lambda$ calculate $\{s_0^\pm, s_1^\pm, c_0^\pm, c_1^\pm\}$ using the results in Sec. II.

(2) Using $\{s_0^\pm, s_1^\pm, c_0^\pm, c_1^\pm\}$ in Eq. (3.9) and Eq. (3.26) calculate $\{\alpha_1^\pm, \beta_1^\pm, \gamma_0^\pm, \rho_0, \sigma_0\}$.

(3) Using $\{\alpha_1^\pm, \beta_1^\pm, \gamma_0^\pm, \rho_0, \sigma_0\}$ in Eq. (A.2) and Eqs. (A.5-A.8) calculate $\{\alpha_2^\pm, \beta_2^\pm, \gamma_1^\pm, \rho_1, \sigma_1\}$.

(4) Using $\{\alpha_2^\pm, \beta_2^\pm, \gamma_1^\pm, \rho_1, \sigma_1\}$ in Eq. (A.2) and Eqs. (A.5-A.8) calculate $\{\alpha_3^\pm, \beta_3^\pm, \gamma_2^\pm, \rho_2, \sigma_2\}$.

(5) Continue until you reach the desired $\{\alpha_{n+1}^\pm, \beta_{n+1}^\pm, \gamma_n^\pm, \rho_n, \sigma_n\}$.

It is important that an effective, stable, and accurate computational routine be employed to obtain the values of the initial set $\{\alpha_1^\pm, \beta_1^\pm, \gamma_0^\pm, \rho_0, \sigma_0\}$ since a large error here will be amplified and propagates to subsequent sets. An advantage of dealing with the ratios $\{\alpha_n^\pm, \beta_n^\pm, \gamma_n^\pm, \rho_n, \sigma_n\}$ rather than the coefficients $\{s_n^\pm, c_n^\pm\}$ is that the exponential $e^{-\mu^2/2}$ that might create calculation problems (e.g., multiplying small values by large ones) at high energies is cancelled out.

## Appendix B: Calculating the matrix elements of the scattering potential $\tilde{V}(x)$ using Gauss quadrature

The matrix elements of $\tilde{V}(x)$ in the basis $\{\xi_m\}_{m=-N}^{N-1} = \{\phi_n^\pm\}_{n=0}^{N-1} \equiv \{\chi_n\}_{n=0}^{2N-1}$ are required (along with those of $H_0$) to calculate the finite 2*N*×2*N* total Hamiltonian matrix $H_0 + \tilde{V}$. These matrix elements are evaluated as the following definite integral

$$\tilde{V}_{nm} = \int_{-\infty}^{+\infty} \chi_n(x)\tilde{V}(x)\chi_m(x)\,dx, \qquad n,m = 0,1,2,..,2N-1, \tag{B.1}$$

that samples the potential along the real line. For a general potential, such an integral is usually done numerically, especially, if $\tilde{V}(x)$ is non-analytic. Using the explicit form of the basis functions in (3.8), we can write this integral as

$$\tilde{V}_{nm} = \int_{-\infty}^{+\infty} \tfrac{1}{\sqrt{\pi}} e^{-y^2} \hat{H}_n(y)\hat{H}_m(y)\tilde{V}(y/\lambda)\,dy, \tag{B.2}$$

where $y = \lambda x$ and $\hat{H}_n(y)$ is the normalized Hermite polynomial, $\hat{H}_n(y) = \frac{1}{\sqrt{2^n\Gamma(n+1)}} H_n(y)$, with the orthogonality relation $\int_{-\infty}^{+\infty} \frac{1}{\sqrt{\pi}} e^{-y^2} \hat{H}_n(y)\hat{H}_m(y)\,dy = \delta_{nm}$ [4]. It satisfies the following symmetric recursion relation

$$y\hat{H}_n(y) = \sqrt{\tfrac{n}{2}}\,\hat{H}_{n-1}(y) + \sqrt{\tfrac{n+1}{2}}\,\hat{H}_{n+1}(y). \tag{B.3}$$

Writing this as the matrix eigenvalue equation $Q|\hat{H}_n\rangle = y|\hat{H}_n\rangle$ gives the following tridiagonal symmetric matrix

$$Q_{n,m} = \sqrt{\tfrac{n}{2}}\,\delta_{n,m+1} + \sqrt{\tfrac{n+1}{2}}\,\delta_{n,m-1}. \tag{B.4}$$

Therefore, Gauss quadrature [13] associated with these normalized Hermite polynomials gives the following approximation for the corresponding integral (B.2)

$$\tilde{V}_{nm} \cong \sum_{k=0}^{K-1} \Omega_{nk} \Omega_{mk} \tilde{V}(\omega_k / \lambda), \tag{B.5}$$

where $\omega_k$ is an eigenvalue of the $K \times K$ tridiagonal matrix $Q$ and $\{\Omega_{nk}\}_{n=0}^{K-1}$ is the corresponding normalized eigenvector. The dimension $K$ of the matrix $Q$ should be large enough to give the desired accuracy for the values of the elements of the scattering potential. It should be greater than $2N$, which is the dimension of the finite total Hamiltonian. We should note that the matrix elements of the potential (B.5) should be re-ordered to match those of $H_0$ for building the full finite Hamiltonian as required by the general construction given by Eq. (3.12).

**Figure captions:**

**Fig. 1:** The partial sums in Eq. (2.23) for the top signs (Fig. 1a) and bottom signs (Fig. 1b). The number of truncated terms $N = 0, 10, 30$.

**Fig. 2:** Schematic diagram for the infinite tridiagonal symmetric matrix representing the reference wave operator (note that $J_{0,-1} = J_{-1,0} = 0$). More details of this matrix in the middle subspace are shown in Fig. 6.

**Fig. 3:** Schematic diagram for the total Hamiltonian. The finite $2N\times 2N$ matrix block on the diagonal is the representation of the total Hamiltonian in the "middle" subspace.

**Fig. 4:** The finite $2N\times 2N$ total Hamiltonian matrix block is made up of four sub-blocks. The two $N\times N$ diagonal sub-blocks represent the even part of the Hamiltonian, whereas, the two $N\times N$ off-diagonal sub-blocks represent the odd part.

**Fig. 5:** A reproduction of Fig. 4 for the case where the left and right ranges of the scattering potential ($N^-$ and $N^+$) are not equal.

**Fig. 6:** Details of the matrix wave equation (3.20).

**Fig. 7:** Reflection and transmission coefficients for the modified hyperbolic Pöschl-Teller potential (4.1) with $\eta$ = 2.0 (a.u.) and $\nu$ = 5/2. The black dashed curves represent exact results.

**Fig. 8:** Reflection and transmission coefficients for the square potential barrier of the second example with $V_0$ = 2.0 (a.u.) and $L$ = 7/2 (a.u.). The black dashed curves represent exact results.

**Fig. 9:** Reflection and transmission coefficients for the double barrier potential (4.6) with $V_0$ = 5.0 (a.u.) and $L$ = 1.0 (a.u.).

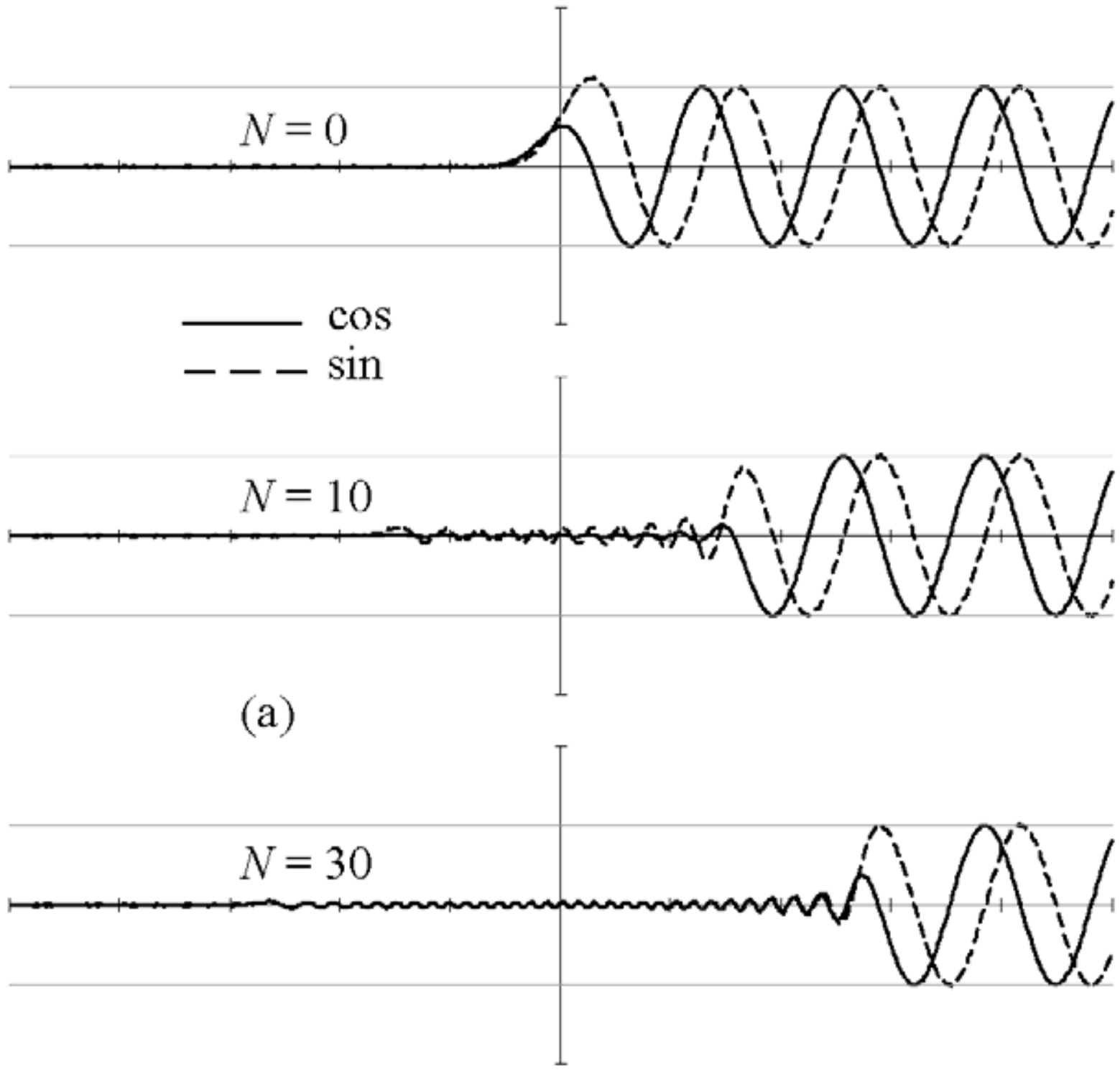

**Fig. 1a**

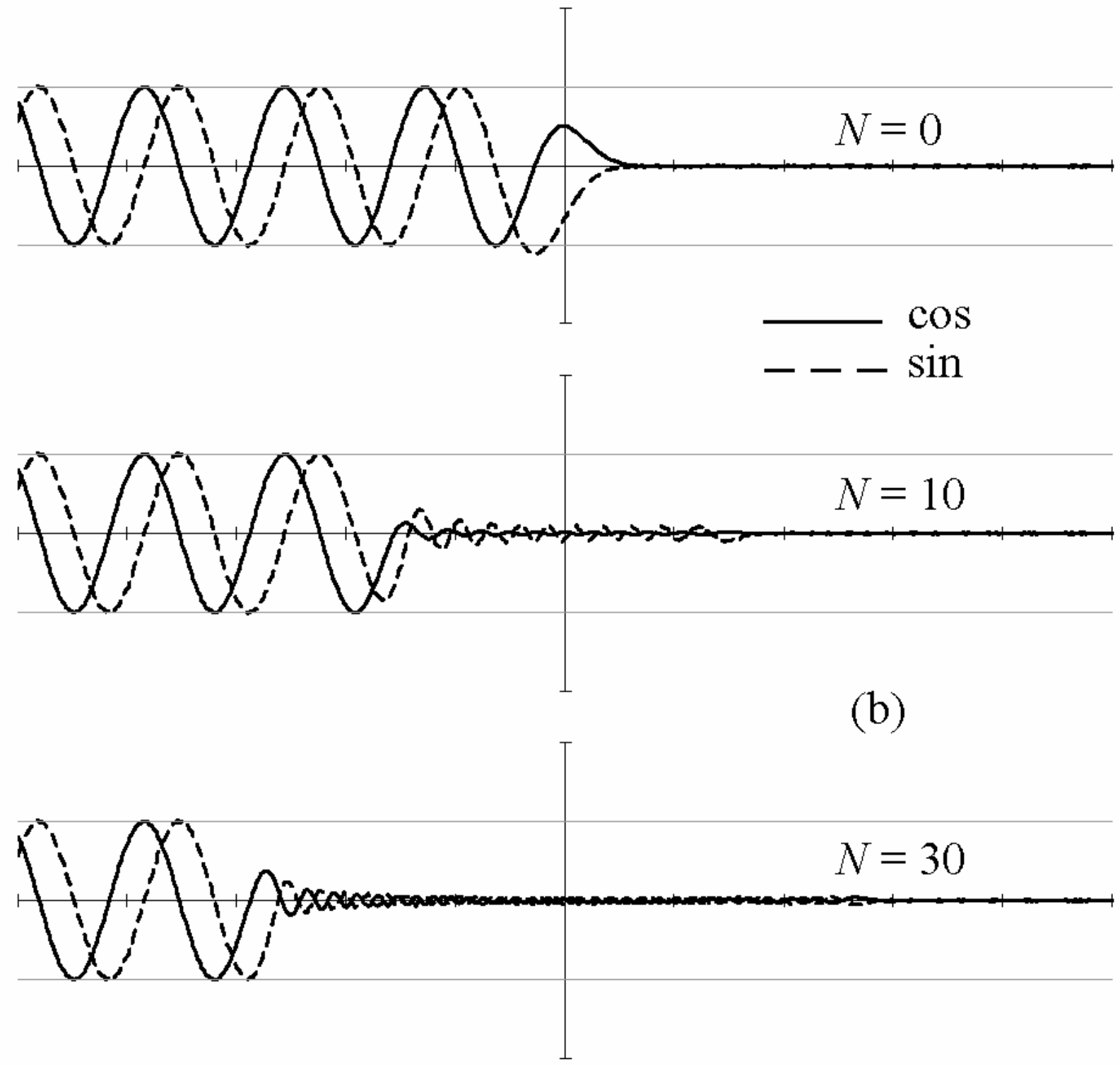

**Fig. 1b**

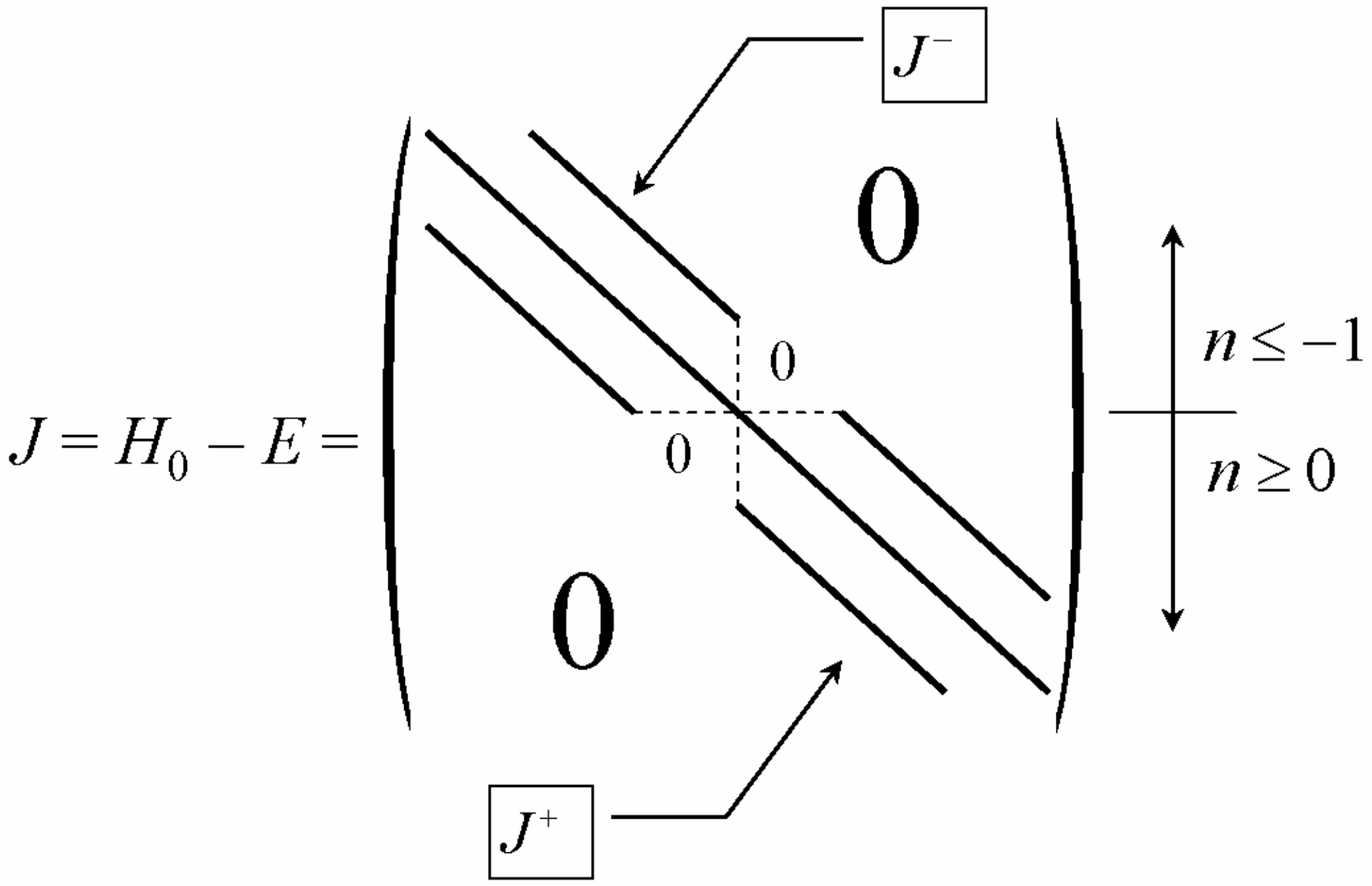

**Fig. 2**

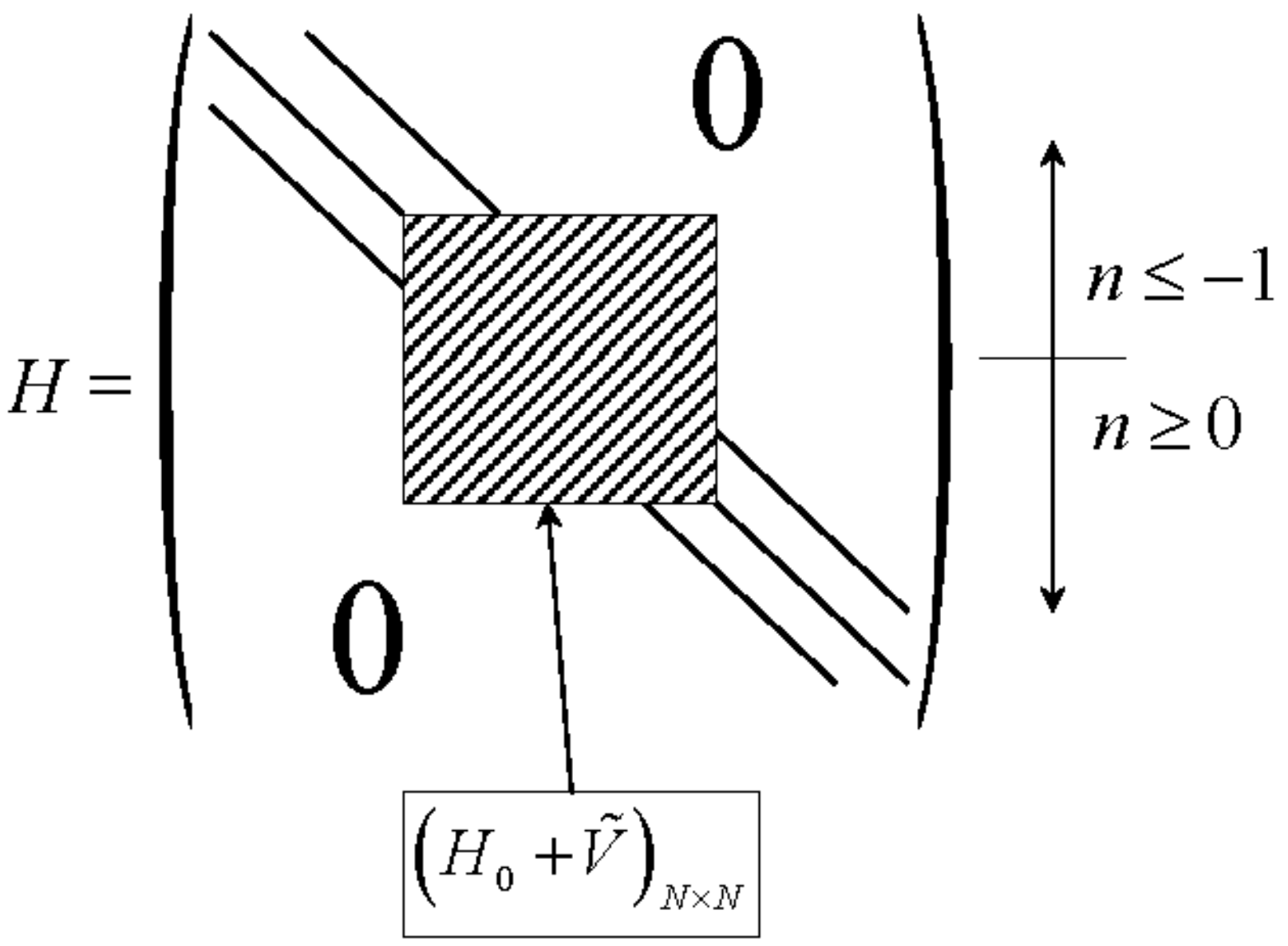

**Fig. 3**

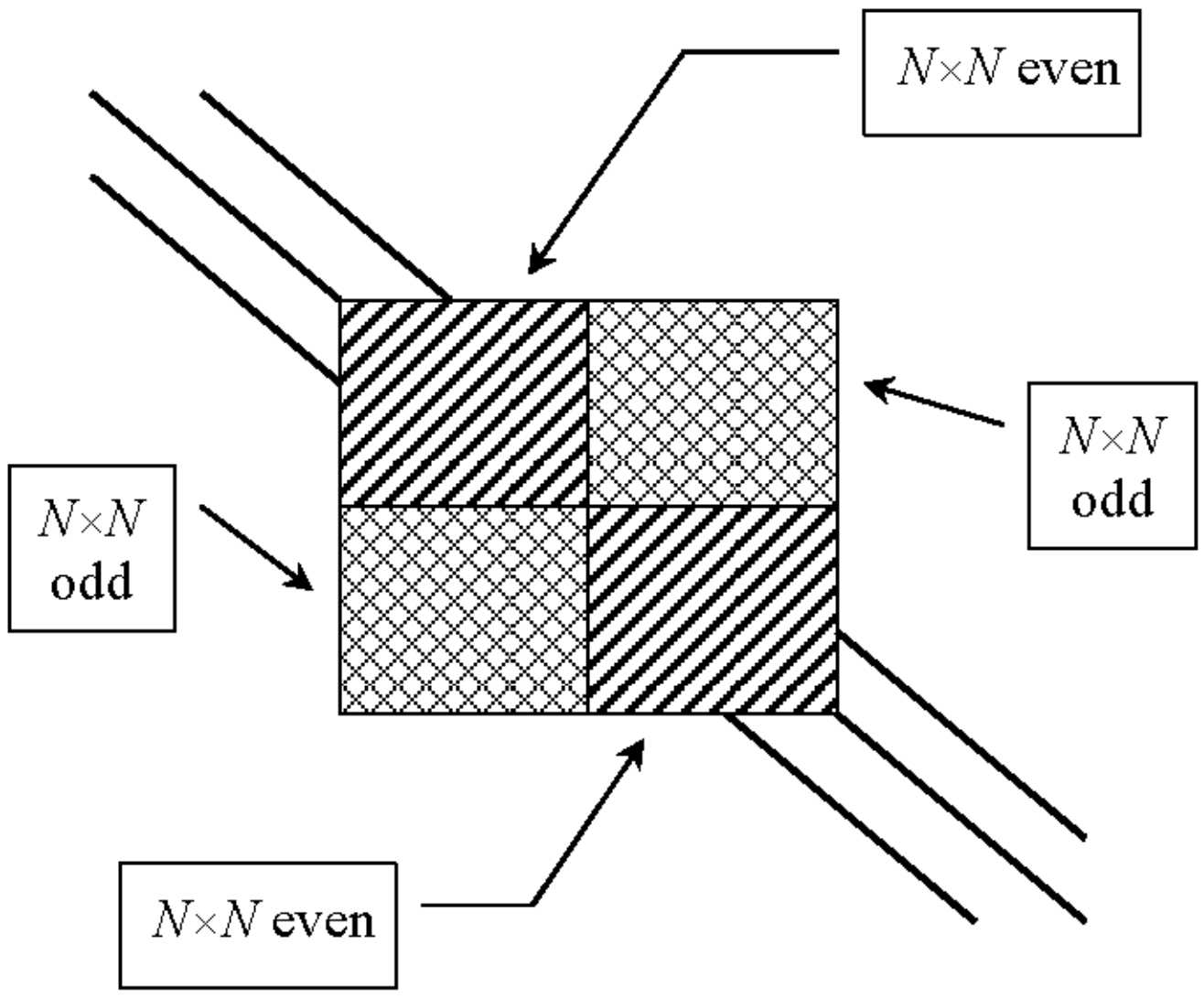

**Fig. 4**

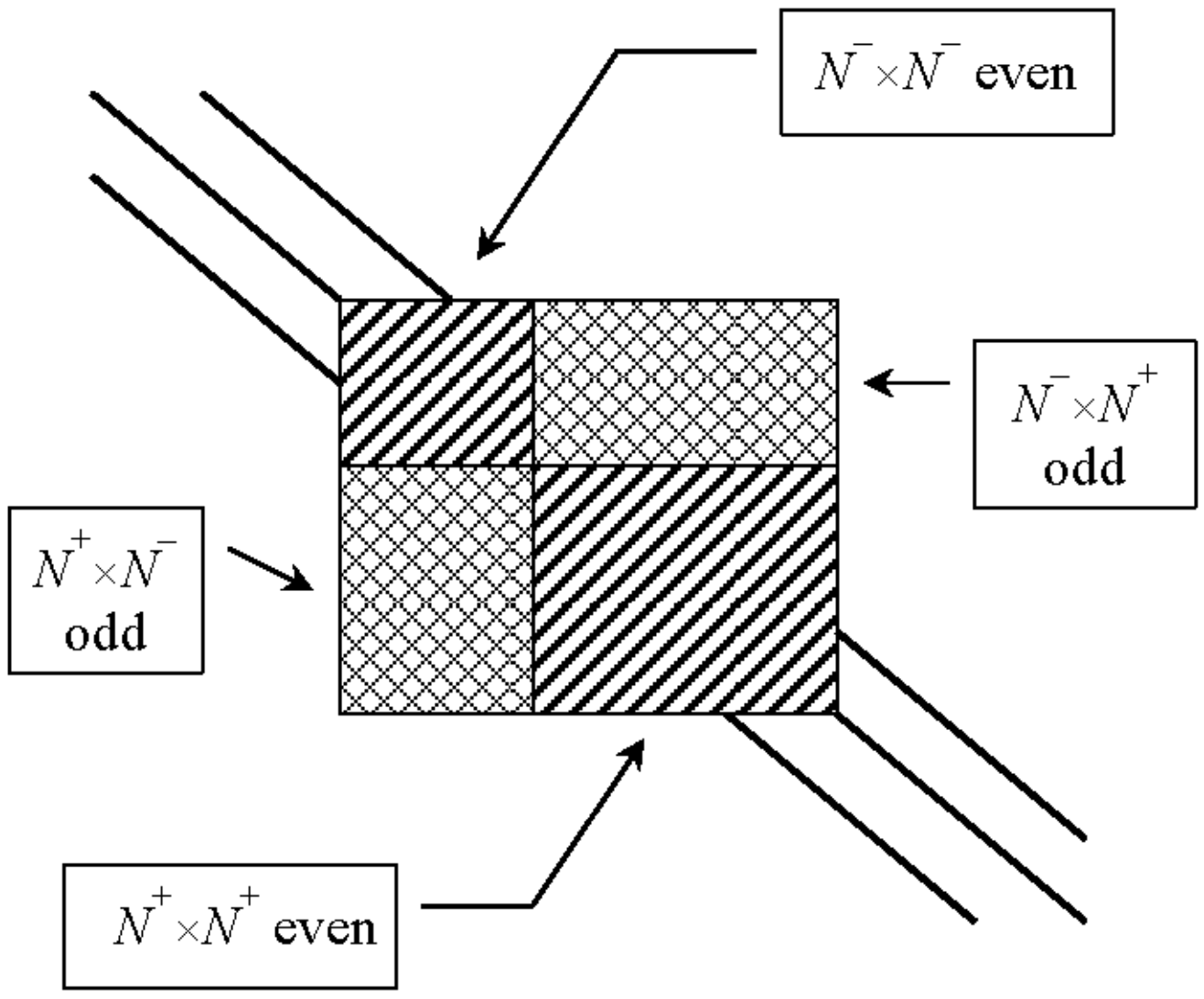

**Fig. 5**

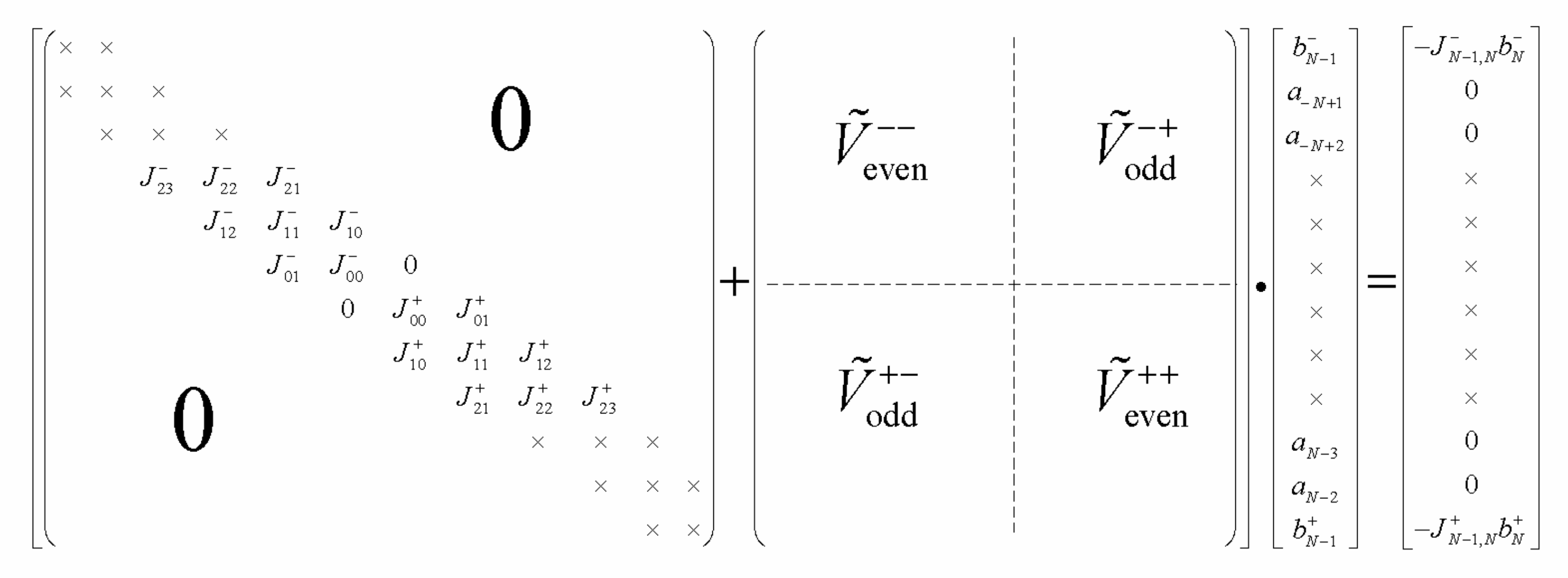

Fig. 6

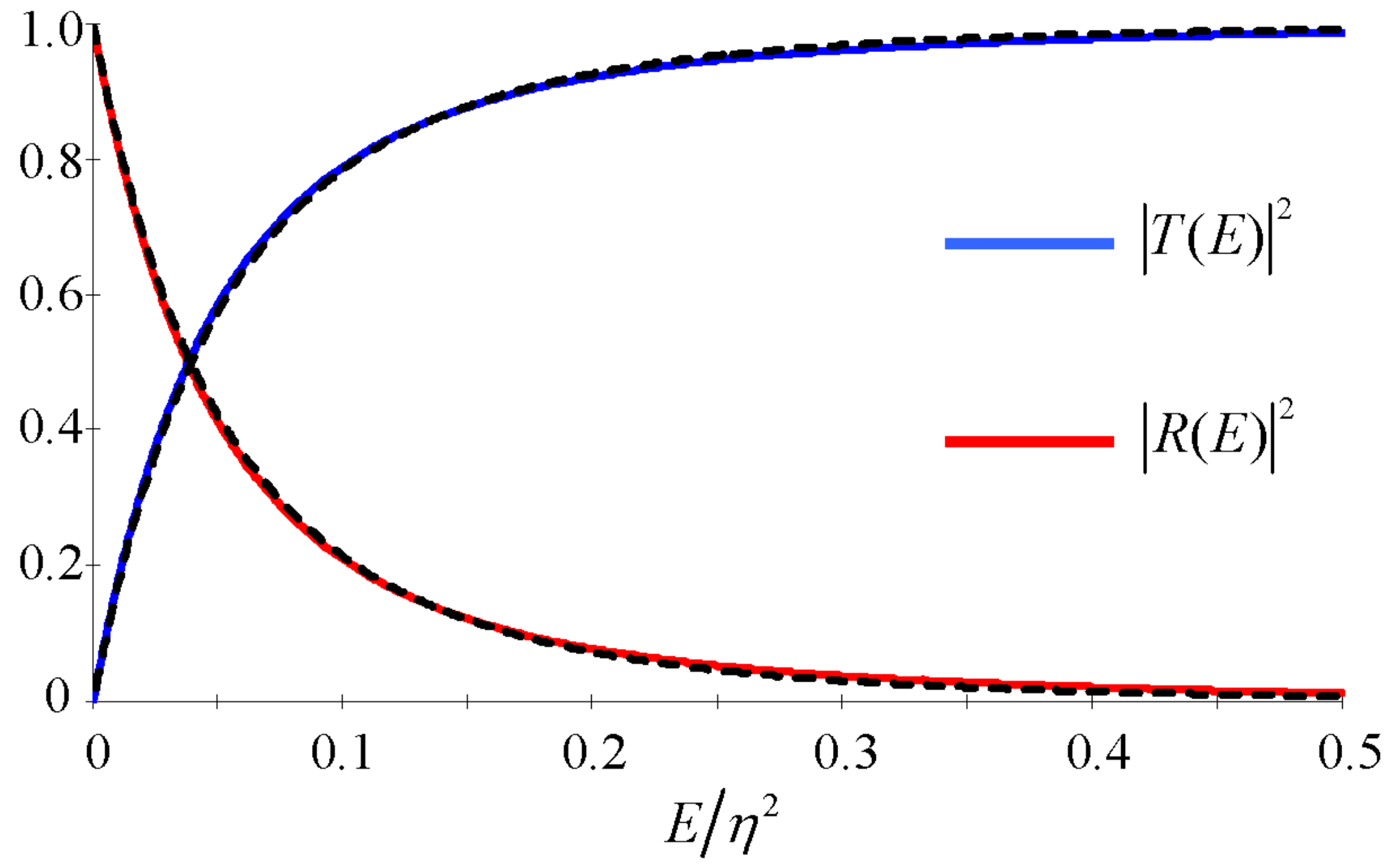

Fig. 7

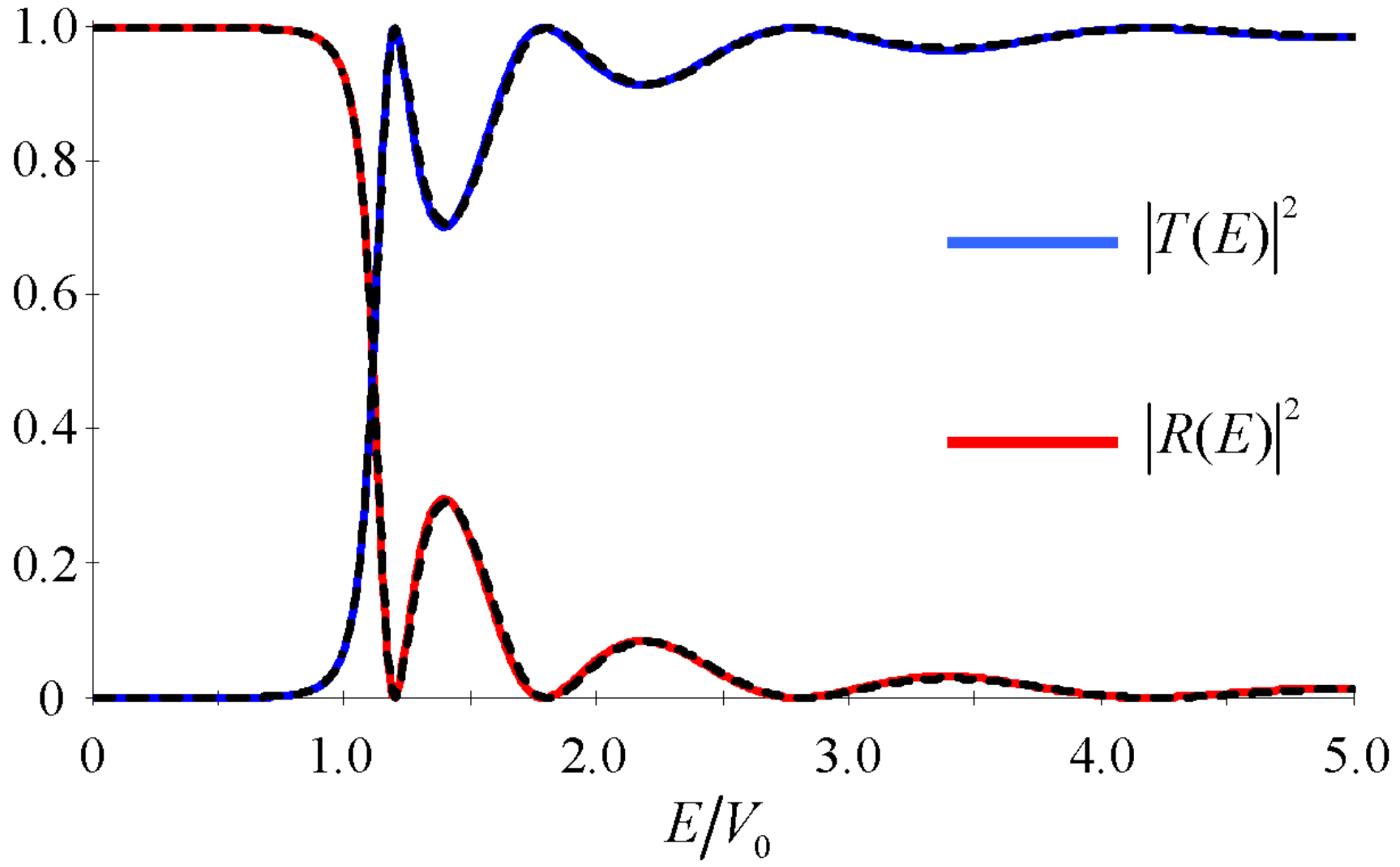

**Fig. 8**

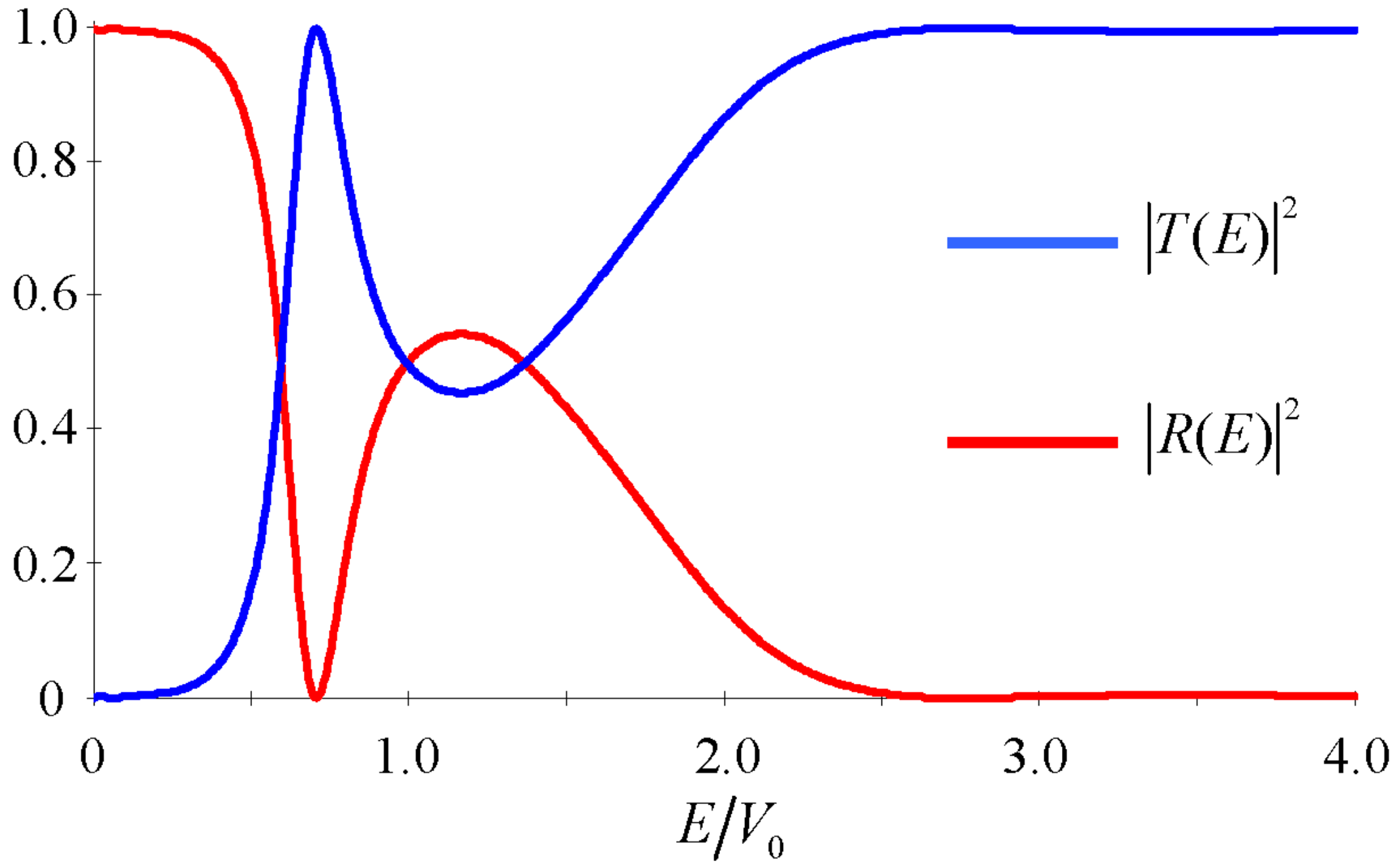

**Fig. 9**